\documentclass[conference,letterpaper]{IEEEtran}

\usepackage{amsmath,amssymb,amsfonts,amsthm}
\usepackage{array,multirow,graphicx}
\usepackage{tikz}
\usetikzlibrary{positioning}
\usepackage{graphicx}
\usepackage{pgfplots} 
\usepackage{mathtools}
\pgfplotsset{compat=newest} 
\usepackage{textcomp}
\usepackage{xcolor}
\usepackage{cite,comment}

\DeclareMathOperator{\EX}{\textrm E}
\DeclareMathOperator{\VarX}{\textrm Var}
\DeclareMathOperator{\trace}{\textrm trace}

\newcommand\figscale{0.85}

\newtheorem{lemma}{Lemma}

\newcommand{\st}{{\tilde s}}

\newcommand{\xt}{{\tilde x}}

\newcommand{\St}{{\tilde S}}

\newcommand{\Xt}{{\tilde X}}
\newcommand{\Yt}{{\tilde Y}}
\newcommand{\Zt}{{\tilde Z}}

\newcommand{\xv}{{\underline x}}

\newcommand{\Sv}{{\underline S}}

\newcommand{\Wv}{{\underline W}}
\newcommand{\Xv}{{\underline X}}
\newcommand{\Yv}{{\underline Y}}
\newcommand{\Zv}{{\underline Z}}

\newcommand{\xtv}{{\underline {\tilde x}}}

\newcommand{\Stv}{{\underline {\tilde S}}}

\newcommand{\Xtv}{{\underline {\tilde X}}}
\newcommand{\Ytv}{{\underline {\tilde Y}}}
\newcommand{\Ztv}{{\underline {\tilde Z}}}

\begin{document}
\title{Achieving Gaussian Vector Broadcast Channel Capacity with Scalar Lattices}
%
\author{%
    \IEEEauthorblockN{M.~Yusuf~\c{S}ener\IEEEauthorrefmark{1}\IEEEauthorrefmark{2}, Gerhard~Kramer\IEEEauthorrefmark{1}, Shlomo Shamai (Shitz)\IEEEauthorrefmark{3}, Ronald B\"ohnke\IEEEauthorrefmark{2}, and Wen Xu\IEEEauthorrefmark{2}}
    \IEEEauthorblockA{\IEEEauthorrefmark{1}%
    School of Computation, Information and Technology, Technical University of Munich, 80333 Munich, Germany}
   \IEEEauthorblockA{\IEEEauthorrefmark{2}%
    Munich Research Center, Huawei Technologies Duesseldorf GmbH, 80992 Munich, Germany}
   \IEEEauthorblockA{\IEEEauthorrefmark{3}%
    Dept. of Electrical and Computer Engineering, Technion—Israel Institute of Technology, Haifa 3200003, Israel}
}

\maketitle
\begin{abstract}
A coding scheme with scalar lattices is applied to K-receiver, Gaussian, vector broadcast channels with K independent messages, one for each receiver. The method decomposes each receiver channel into parallel scalar channels with known interference and applies dirty paper coding with a modulo interval, amplitude shift keying (ASK), and probabilistic shaping to each scalar channel. The achievable rate tuples include all points inside the capacity region by choosing truncated Gaussian shaping, large ASK alphabets, and large modulo intervals.
\end{abstract}


\section{Introduction}
Dirty paper coding (DPC) with scalar lattices can achieve the capacity of the dirty paper channel \cite{csener2024-dpc-capacity}; cf. \cite{csener2021dirty,senerdpcmiso}. The result suggests that a similar scheme can achieve the capacity of multi-input, multi-output (MIMO) broadcast channels, and the purpose of this paper is to prove this. We do so in two steps. First, we apply noise whitening and the singular value decomposition (SVD) to each receiver channel to obtain parallel scalar channels with known interference. Second, we extend the theory in \cite{csener2024-dpc-capacity} to show that scalar DPC with $M$-ary amplitude shift keying (ASK), a modulo operator with interval length $A$, and truncated Gaussian shaping can achieve any rate tuple inside the capacity region of $K$-receiver, Gaussian MIMO broadcast channels with $K$ independent messages, one for each receiver, by choosing large $M$ and $A$.

This paper is organized as follows. Sec.~\ref{sec:preliminaries} reviews notation and results on symmetric unimodal functions. Sec.~\ref{sec:MIMO-BC} describes the model and decomposes the vector broadcast channel into independent parallel channels to which we apply scalar DPC. Sec.~\ref{sec:lemma-2} proves Lemma~2 in \cite{csener2024-dpc-capacity}. Sec.~\ref{sec:theorem-1} proves an extension of Theorem~1 in \cite{csener2024-dpc-capacity}. Sec.~\ref{sec:conc} concludes the paper.

\section{Preliminaries}
\label{sec:preliminaries}
%
\subsection{Notation}
Underlined letters such as $\xv=[x_1,\dots,x_n]^T$ refer to column vectors, where $\xv^T$ is the transpose of $\xv$. Bold letters such as ${\bf Q}$ denote matrices; ${\bf Q}^\dag$ is the complex-conjugate transpose of ${\bf Q}$; $\trace({\bf Q})$ is the trace of ${\bf Q}$. We write $j=\sqrt{-1}$.

Upper and lowercase letters refer to random variables (RVs) and vectors and their realizations, e.g., $\Xv$ is a random vector and $\xv$ is its realization. We write $P_X$ and $p_X$ for probability mass functions and densities, respectively. We remove subscripts if the argument is a lowercase version of the RV, e.g., $P(x)=P_X(x)$. $\EX[X]$ and $\VarX[X]$ are the expectation and variance of $X$, respectively. $h(X)$ is the differential entropy of $X$ and $I(X;Y)$ is the mutual information of $X$ and $Y$. We use natural logarithms. We write $[x]^+=\max(0,x)$ and
\begin{align}
    x\ \textrm{mod}\ A = x - k' A
\end{align}
where $k'$ is the integer so that $x-k' A$ lies in $[-A/2,A/2)$.

\subsection{Symmetric and Unimodal Functions}
The real-valued function $f(.)$ is called symmetric if $f(x)=f(-x)$ for all $x\in\mathbb R$.
We prove the following (known) lemma in the Appendix.

\begin{lemma} \label{lemma:symmetric}
    The convolution $f*g(.)$ of two symmetric functions $f(.)$ and $g(.)$ is symmetric.
\end{lemma}

We will study symmetric and unimodal probability density functions (p.d.f.s) $p(.)$, i.e., $p(x)$ is non-increasing for $x\ge0$. The following lemma was proved in \cite[pp. 30-32]{wintner-38} and \cite[Thm.~2.1]{Purkayastha-98}; we provide an alternative proof in the Appendix.

\begin{lemma} \label{lemma:symmetric-unimodal-conv}
    The convolution $p*q(.)$ of two symmetric unimodal p.d.f.s $p(.)$ and $q(.)$ is symmetric unimodal.
\end{lemma}

Finally, we derive bounds on the sum of uniformly spaced samples of a symmetric unimodal function $f(.)$. We prove the following lemma in the Appendix for the spacing $A/M$. 

\begin{lemma} \label{lemma:symmetric-unimodal-sum}
    Consider a symmetric unimodal $f(.)$. We have
    \begin{align}
        \left| \int_{-\frac{A}{2}}^{\frac{A}{2}} f(y) dy -
        \sum_{k=-\lfloor M/2 \rfloor}^{\lceil M/2 \rceil-1} \frac{A}{M} f\left(x + k \frac{A}{M} \right) \right| \le \frac{A}{M} f(0) \label{eq:sum-bound} 
    \end{align}
    for $x$ satisfying $0\le x< A/(2M)$. Similarly, if $f(.)$ has finite area then for any $x$ we have
    \begin{align}
        \left| \int_{\mathbb R} f(y) dy -
        \sum_{k\in\mathbb Z} \frac{A}{M} f\left(x + k \frac{A}{M} \right) \right| \le \frac{A}{M} f(0) . \label{eq:sum-bound2} 
    \end{align}
\end{lemma}

\section{Gaussian Vector Broadcast Channels}
\label{sec:MIMO-BC}
%
\subsection{Model and Coding Schemes}
\label{subsec:model-coding}
Consider complex-alphabet channels with a $n_\textrm{t}$-dimensional input $\Xv$ and
$n_k$-dimensional outputs
\begin{align}
    \Yv_k = {\bf H}_k \Xv + \Zv_k, \quad k=1,\dots,K
    \label{eq:MIMO-BC}
\end{align}
where ${\bf H}_k$ is a $n_k \times n_\textrm{t}$ complex matrix, $\Zv_k$ is $n_k$-dimensional, circularly-symmetric, complex, Gaussian (CSCG) noise with invertible covariance matrix ${\bf Q}_k$, and $\Xv$ satisfies the power constraint $\EX[\|\Xv\|^2]\le P_X$.
It is known that DPC with typical-sequence binning \cite{marton-IT79,gelfand1980coding,costa1983writing,Caire-Shamai-IT03,vishwanath2003duality,weingarten2006capacity,Geng-Nair-IT14} or high-dimensional lattice coding \cite{erez2005capacity,erez2005close,bennatan2006superposition,sun2008nested,sun2009joint,shilpa2010dirty} achieves capacity if there is a dedicated message for each receiver. However, the complexity of these schemes is prohibitive. Simpler lattice schemes with integer-forcing appear in \cite{gariby07,Silva-WCOM17,He-IT18,Venturelli-WCOMMLETT20}; these perform well but do not approach capacity in general. Concrete codes are described in \cite{erez2005close,sun2008nested,sun2009joint,shilpa2010dirty,bennatan2006superposition,Korada-IT10,liu2016polar,csener2021dirty}. Non-asymptotic
analyses with random codes are provided in \cite{Liu-IT06,Verdu-Allerton12,Watanabe-IT15,Scarlett-IT15,senerdpcmiso,csener2024-dpc-capacity,Tamir-IT23}.

\subsection{Capacity-Achieving Scheme}
\label{subsec:capacity-achieving-scheme}
The capacity-achieving scheme performs successive DPC at the transmitter for all receiver orderings. We describe the approach for the ordering $1,\dots,K$. The transmitter sends $\Xv=\sum_{k=1}^K \Xv_k$ where the $\Xv_k$ are statistically independent \cite{vishwanath2003duality} and each $\Xv_k$ has a covariance matrix ${\bf K}_k$ optimized for a particular rate tuple. The optimal $\Xv_k$ are CSCG.

Receiver $k$ sees
\begin{align}
    \Yv_k
    & = {\bf H}_k \Xv_k + \underbrace{\left( \sum_{l<k} {\bf H}_k \Xv_l \right)}_{\textstyle :=\Sv_k} + \underbrace{\left( \sum_{l>k} {\bf H}_k \Xv_l \right) + \Zv_k}_{\textstyle :=\underline {\check Z}_k}
    \label{eq:Yk}
\end{align}
and treats $\Sv_k$ as interference and $\underline {\check Z}_k$ as noise. Let ${\bf \check Q}_k$ be the covariance matrix of $\underline {\check Z}_k$. One may use the SVD to write
\begin{align}
    {\bf \check Q}_k^{-1/2} {\bf H}_k {\bf K}_k^{1/2}
    = {\bf U}_k {\bf \Sigma}_k {\bf V}_k^\dag
    \label{eq:SVD}
\end{align}
where the ${\bf U}_k$ and ${\bf V}_k$ are unitary matrices and ${\bf \Sigma}_k$ is a $n_k \times n_\textrm{t}$ diagonal matrix with the singular values of ${\bf \check Q}_k^{-1/2} {\bf H}_k {\bf K}_k^{1/2}$. Receiver $k$ left-multiplies $\Yv_k$ with the noise-whitening filter ${\bf U}_k^\dag {\bf \check Q}_k^{-1/2}$ to obtain the parallel dirty paper channel
\begin{align}
    \Ytv_k = {\bf \Sigma}_k \Xtv_k + \Stv_k + \Ztv_k
    \label{eq:Ytvk}
\end{align}
where $\Stv_k$ is known at the transmitter and
\begin{align}
    \Xv_k  & =  {\bf K}_k^{1/2} {\bf V}_k \Xtv_k \label{eq:Xtvk} \\
    \Stv_k & = {\bf U}_k^\dag {\bf \check Q}_k^{-1/2} \Sv_k \label{eq:Stvk} \\
    \Ztv_k & = {\bf U}_k^\dag {\bf \check Q}_k^{-1/2} \underline {\check Z}_k.
    \label{eq:Ztvk}
\end{align}
The covariance matrices of $\Xtv_k$ and $\Ztv_k$ are identity matrices, i.e., the entries $\Xt_{k,i}$, $i=1,\dots,n_\textrm{t}$, and $\Zt_{k,i}$, $i=1,\dots,n_k$, are independent and identically distributed (i.i.d.) CSCG with unit variance \cite{Neeser-IT93} (we write the form \eqref{eq:Xtvk} since ${\bf K}_k$ could be singular; in this case more choices are possible for $\Xtv_k$). The power constraint requires $\sum_{k=1}^K P_k \le P_X$ where $P_k=\trace({\bf K}_k)$.

\subsection{Parallel DPC with Scalar Lattices}
\label{subsec:parallel-DPC}
Observe from \eqref{eq:Xtvk} that the new vectors $\Xtv_k$, $k=1,\dots,K$, are independent since the $\Xv_k$ are independent. Moreover, we can write the $i$th entry of $\Ytv_k$ in \eqref{eq:Ytvk} as
\begin{align}
    \Yt_{k,i} = \sigma_{k,i} \Xt_{k,i} + \St_{k,i} + \Zt_{k,i}
    \label{eq:Ytki}
\end{align}
where $\sigma_{k,i}$ is a singular value, $\St_{k,i}$ is known at the transmitter, and $\Zt_{k,i}$ is CSCG and independent of $\Xt_{k,i},\St_{k,i}$. Thus, one achieves the capacity of MIMO broadcast channels if one achieves the capacity $\log(1+\sigma_{k,i}^2)$ for each scalar dirty paper channel~\eqref{eq:Ytki} in~\eqref{eq:Ytvk}, and for the covariance matrices ${\bf K}_1,\dots,{\bf K}_K$ of all required rate tuples.

We may write $\Zv_k={\bf Q}_k^{1/2} \Wv_k$ where the entries $W_{k,i}$ of $\Wv_k$ are i.i.d. CSCG with unit variance. Thus, using \eqref{eq:Yk}, \eqref{eq:Xtvk}, and \eqref{eq:Ztvk}, we have
\begin{align}
    \Ztv_k = {\bf U}_k^\dag {\bf \check Q}_k^{-1/2} \left[ \left(\sum_{l>k} {\bf H}_k  {\bf K}_l^{1/2} {\bf V}_l \Xtv_l\right)+{\bf Q}_k^{1/2} \Wv_k\right].
\end{align}
Alternatively, there are constants $a_{k,i,l,h}$ and $b_{k,i,h}$ for which
\begin{align}
    \Zt_{k,i} = \left(\sum_{l>k} \sum_{h=1}^{n_\textrm{t}} a_{k,i,l,h} \Xt_{l,h}\right) + \sum_{h=1}^{n_k} b_{k,i,h} W_{k,h}.
    \label{eq:Ztki}
\end{align}
We now apply the DPC method in \cite{csener2021dirty} to each channel \eqref{eq:Ytki} by treating the real and imaginary parts of $\Yt_{k,i}$ as independent with the same channel gain $\sigma_{k,i}$ and noise variance $1/2$. We focus on the real part and abuse notation by using the same symbols as for the complex alphabet channels, i.e., all random variables are now real-valued. Also, the $\Xt_{k,i}$, $\St_{k,i}$, and $\Zt_{k,i}$ are no longer Gaussian, with the exception of the $\Zt_{K,i}$.

More precisely, consider real-alphabet channels~\eqref{eq:Ytki} and symbols $U_{k,i}$ with $M_{k,i}$-ASK alphabets $\mathcal{U}_{k,i}$ for all $k$, $i$. The encoder computes (see \cite[eq.~(1)-(2)]{csener2024-dpc-capacity})
\begin{align}
    \St_{k,i}' & = (\alpha_{k,i} \St_{k,i} + D_{k,i})\ \textrm{mod}\ A_{k,i} \\
    \Xt_{k,i} & = (U_{k,i} - \St_{k,i}')\ \textrm{mod}\ A_{k,i} \label{eq:xtki}
\end{align}
where $U_{k,i}$ is selected using $\St_{k,i}'$ and a shaping density $q_{k,i}(.)$ (see \cite[eq.~(7)]{csener2024-dpc-capacity}), $\alpha_{k,i}$ is a minimum mean square error (MMSE) coefficient, and the dither $D_{k,i}$ is continuously uniform over the modulo interval $[-A_{k,i}/2,A_{k,i}/2)$.

We choose the $D_{k,i}$ as mutually independent, which makes the $\Xt_{k,i}$ mutually independent for all $k,i$; see the Appendix. Thus, all RVs on the right-hand side of \eqref{eq:Ztki} are independent and the density of $\Zt_{k,i}$ is the convolution of the densities of the $a_{k,i,l,h} \Xt_{l,h}$ and $b_{k,i,h} W_{k,h}$ (more generally, if a $\Xt_{l,h}$ does not have a density, the density of $\Zt_{k,i}$ is the derivative of the convolution of the distribution functions of the $a_{k,i,l,h} \Xt_{l,h}$ and $b_{k,i,h} W_{k,h}$). Compared to \cite{csener2024-dpc-capacity}, the main change is that $\Zt_{k,i}$ is not necessarily Gaussian. However, we argue that all steps in \cite[eq.~(9)-(18)]{csener2024-dpc-capacity} remain valid.
\begin{itemize}
    \item The $Z'$ in \cite[eq.~(4)]{csener2024-dpc-capacity} is now
\begin{align}
    Z' := Z_{k,i}' = (\alpha_{k,i}-1) \Xt_{k,i} + \alpha_{k,i} \Zt_{k,i}
    \label{eq:Z'}
\end{align}
    where $\Xt_{k,i}$ and $\Zt_{k,i}$ are independent and $\alpha_{k,i}=P_X/(P_X+P_Z)$ with $P_X=\EX[\Xt_{k,i}^2]$ and $P_Z=\EX[\Zt_{k,i}^2]$.
    \item The dither induces a discretely uniform $U:=U_{k,i}$, and the same steps as in the proof of Lemma~1 in \cite{csener2024-dpc-capacity} show that $U_{k,i}$ and $(\Xt_{k,i},Z_{k,i}')$ are independent.
    \item Lemma~2 in \cite{csener2024-dpc-capacity} remains valid because it depends on the encoding only. We prove this Lemma in Sec.~\ref{sec:lemma-2} below.
    \item Theorem~1 in \cite{csener2024-dpc-capacity} remains valid; see Sec.~\ref{sec:theorem-1} below. The main change compared to \cite{csener2024-dpc-capacity} is that $Z_{k,i}'$ in \eqref{eq:Z'} is now a mixture of several Gaussian and shaped RVs rather than one Gaussian and one shaped RV.
\end{itemize}

\section{Proof of Lemma~2 in \cite{csener2024-dpc-capacity}}
\label{sec:lemma-2}

For any density $q(.)$, consider the function (see~\cite[eq.~(8)]{csener2024-dpc-capacity})
\begin{align}
    d(x) = \sum_{k=0}^{M-1} \frac{A}{M}
    q\left( (x+k\frac{A}{M}\right) \textrm{mod}\ A).
    \label{eq:d}
\end{align}
Lemma~1 in \cite{csener2024-dpc-capacity} states that the density of the dirty paper channel input $X$ is $p(x)=q(x)/d(x)$ for $x\in[-A/2,A/2)$. The proof of \cite[Lemma~2]{csener2024-dpc-capacity} argues directly that the Riemann sum \eqref{eq:d} satisfies $\lim_{M\rightarrow\infty} d(x)=1$. Here, we consider finite $M$ and obtain more insight.

Observe that for $x\ \textrm{mod}\ A/M = x-k'(A/M)$, we have
\begin{align}
    d(x) & =
    \sum\nolimits_{k=0}^{M-1} \frac{A}{M}  q\left(\left(x + (k+k')\frac{A}{M}\right)\ \textrm{mod}\ A\right) \nonumber \\
    & \overset{(a)}{=} 
    \sum\nolimits_{k=0}^{M-1} \frac{A}{M}  q\left(\left(x\ \textrm{mod}\ \frac{A}{M} + k\frac{A}{M}\right)\ \textrm{mod}\ A\right) \nonumber \\
    & = d\left(x\ \textrm{mod}\ \frac{A}{M}\right)
    \label{eq:dx}
\end{align}
where step $(a)$ follows because the modulo operator permits summing over any $M$ successive integers. Thus, $d(.)$ is periodic with period $A/M$.
Next, suppose $q(.)$ is symmetric; this implies $d(.)$ and $p(.)$ are symmetric. Moreover, we may focus on $x\in[0,A/(2M))$ for which we can write
\begin{align}
    d(x) & = \sum_{k=-\lfloor M/2 \rfloor}^{k=\lceil M/2 \rceil-1} \frac{A}{M} q\left(x+k\frac{A}{M}\right).
    \label{eq:D-sum}
\end{align}

\begin{figure}[!t]
\centering
\scalebox{\figscale}{\input{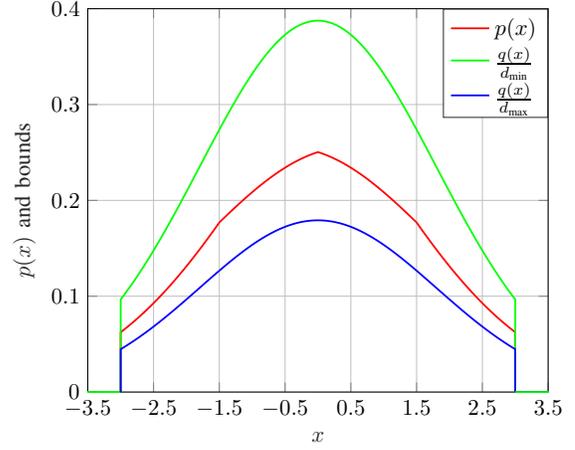}}
\caption{Curves of \eqref{eq:p-bounds} for $A=6$, $M=4$, and truncated Gaussian shaping.}
\label{fig:bounds}
\end{figure}

Finally, suppose $q(.)$ is symmetric unimodal. Applying \eqref{eq:sum-bound} in Lemma~\ref{lemma:symmetric-unimodal-sum}, we have $d_\textrm{min} \le d(x) \le d_\textrm{max}$ for all $x$, where
\begin{align}
    d_\textrm{min} = 1 - \frac{A}{M} q(0), \quad 
    d_\textrm{max} = 1 + \frac{A}{M} q(0).
\label{eq:D-extrema}
\end{align}
We thus have $d_\textrm{min},d_\textrm{max}\rightarrow 1$ for $M\rightarrow\infty$ and
\begin{align}
    \frac{q(x)}{d_\textrm{max}} \le p(x) \le \frac{q(x)}{d_\textrm{min}}
    \label{eq:p-bounds}
\end{align}
where we assume $M$ is sufficiently large so $d_\textrm{min}>0$. Fig.~\ref{fig:bounds} illustrates the bounds \eqref{eq:p-bounds} for $A=6$, $M=4$, and the truncated Gaussian $q(.)$ in \cite[Eq.~(12)]{csener2024-dpc-capacity} with $\sigma_X=1.8$. We compute $d_\textrm{min}\approx 0.63$ and $d_\textrm{max}\approx 1.37$. The bounds are loose because $M$ is relatively small.

\section{Extension of Theorem~1 in \cite{csener2024-dpc-capacity}}
\label{sec:theorem-1}

We generalize the proof of \cite[Theorem~1]{csener2024-dpc-capacity} to include finite $M$ and noise $Z'$ that is a sum of several Gaussian and shaped RVs, as in~\eqref{eq:Z'}, rather than one Gaussian and one shaped RV, as in \cite[eq.~(4)]{csener2024-dpc-capacity}. We first bound the $P_X$ and $h(X)$ in \cite[eq.~(15)--(17)]{csener2024-dpc-capacity} and then study the $Y'$ in \cite[eq.~(4)]{csener2024-dpc-capacity}.

\subsection{Power and Differential Entropy of $\Xt_{k,i}$}
\label{subsec:power-X}

Observe that $X$ in \cite[eq.~(2)]{csener2024-dpc-capacity} plays the role of $\Xt_{k,i}$ here, so define $X:=\Xt_{k,i}$. The power $P_X=\EX[X^2]$ is based on an expectation with respect to $p(.)$. For convenience, define the expectation with respect to $q(.)$ as
\begin{align}
    \EX_q[f(X)] := \int_\mathbb{R} q(x) f(x) dx.
    \label{eq:EXq}
\end{align}
Using \eqref{eq:p-bounds}, for symmetric unimodal $q(.)$, we have
\begin{align}
\frac{\EX_q[X^2]}{d_\textrm{max}} \le P_X \le \frac{\EX_q[X^2]}{d_\textrm{min}}
\label{eq:power_relation}
\end{align}
which proves \cite[eq.~(16)]{csener2024-dpc-capacity} for the truncated Gaussian $q(.)$ in \cite[Eq.~(12)]{csener2024-dpc-capacity} and $M\rightarrow\infty$.
Similar to \eqref{eq:EXq}, define
\begin{align}
    h_q(X) := \EX_q[-\log q(X)].
\end{align}
Using \eqref{eq:p-bounds}, we have (see the Appendix)
\begin{align}
    h(X) \ge {\displaystyle \frac{h_q(X)}{d_\textrm{max}}} + \log d_\textrm{min} - \!\left(\frac{1}{d_\textrm{min}}-\frac{1}{d_\textrm{max}} \right) \! \left[q(0) \log q(0) \right]^+ 
    \label{eq:entropy-bound-1} \\
    h(X) \le {\displaystyle \frac{h_q(X)}{d_\textrm{min}}} + \log d_\textrm{max} + \!\left(\frac{1}{d_\textrm{min}} - \frac{1}{d_\textrm{max}} \right) \! \left[q(0) \log q(0) \right]^+
    \label{eq:entropy-bound-2}
\end{align}
which proves \cite[eq.~(17)]{csener2024-dpc-capacity} for $M\rightarrow\infty$.

\subsection{Density and Differential Entropy of $Y'$}
\label{subsec:density-Y'}

We bound the density and entropy of $Y'$ in \cite[eq.~(4)]{csener2024-dpc-capacity} but where $Z'$ is given by~\eqref{eq:Z'}. Note that $\Zt_{k,i}$ in~\eqref{eq:Ztki} replaces $Z$ in \cite[eq.~(4)]{csener2024-dpc-capacity}.
Consider discretely uniform $M$-ary $U_{k,i}$ and
\begin{align}
 Y' := Y'_{k,i} = (U_{k,i} + Z'_{k,i})\ \textrm{mod}\ A
 \label{eq:eff_chn}
\end{align}
where $U_{k,i}$ and $Z'_{k,i}$ are independent. Suppose $M$ is even; the case where $M$ is odd can be treated similarly. Using $Z'=Z'_{k,i}$ as in~\eqref{eq:Z'}, we have $p(z')=p(-z')$ and
\begin{align}
    p(y') & = \sum_{\ell\in\mathbb{Z}} \sum_{v\in\mathcal{U}_{k,i}} \frac{1}{M} p_{Z'}(y'-v-\ell A) \nonumber \\
    & = \frac{1}{A} \sum_{k \in\mathbb{Z}} \frac{A}{M} p_{Z'}\left(y' + k \frac{A}{M} + \frac{A}{2M} \right) \nonumber \\
    & = p_{Y'}\left( y'\ \textrm{mod} \frac{A}{M} \right)
    \label{eq:pyp}
\end{align}
for $y' \in [-A/2,A/2)$ and $p(y')=0$ otherwise. Thus, $p(y')$ is symmetric and circularly periodic with period $A/M$. 

Next, using \eqref{eq:Ztki} and \eqref{eq:Z'}, the density $p(z')$ is the convolution of three classes of densities: $(\alpha_{k,i}-1)\Xt_{k,i}$, the $\alpha_{k,i}\, a_{k,i,l,h} \Xt_{l,h}$ for $l>k$ and all $h$, and the $\alpha_{k,i}\, b_{k,i,h} W_{k,h}$ for all $h$. The densities of $\Xt_{k,i}$ and the $\Xt_{l,h}$ are chosen to be symmetric unimodal and can each be lower-bounded by an appropriate $q(.)/d_\textrm{max}$ as in~\eqref{eq:p-bounds}. Moreover, the Gaussian densities of the $W_{k,h}$ are also symmetric unimodal. Thus, by Lemma~\ref{lemma:symmetric-unimodal-conv} and using~\eqref{eq:D-extrema}--\eqref{eq:p-bounds}, we can lower bound $p_{Z'}(.)$ by a symmetric unimodal $\underline{p}(.)$ that approaches $p_{Z'}(.)$ for large $M$. Similarly, we can upper bound $p_{Z'}(.)$ by a symmetric unimodal $\overline{p}(.)$ that approaches $p_{Z'}(.)$ for large $M$.

For example, the dirty paper channel in \cite{csener2024-dpc-capacity} corresponds to having $K=1$ receiver, and we compute
\begin{align}
    p(z') & = p_{(\alpha-1)X}*p_{\alpha Z}(z') \nonumber \\
    & \overset{(a)}{=} \int_{-(1-\alpha)A/2}^{(1-\alpha)A/2} \frac{p_X\left(\tilde x/(\alpha-1)\right)}{1-\alpha}\, p_{\alpha Z}(z'-\tilde x)\, d\tilde x \nonumber \\
    & \overset{(b)}{\ge} \underbrace{\int_{-A/2}^{A/2} \frac{q(x)}{d_\textrm{max}}\, p_{\alpha Z}\left(z'+x(1-\alpha)\right)\, dx}_{\textstyle := \underline{p}(z')}
    \label{eq:pzp-lower}
\end{align}
where step $(a)$ follows by $p_{cX}(\tilde x)=p_X(\tilde x/c)/|c|$ for $c\ne0$, and step $(b)$ follows by \eqref{eq:p-bounds} and substituting $x=\tilde x/(\alpha-1)$. Let
$\overline{p}(z')$ be the same as $\underline{p}(z')$ but with $d_\textrm{min}$ replacing $d_\textrm{max}$ in~\eqref{eq:pzp-lower}. Fig.~\ref{fig:pz} shows $p(z')$ and the bounds $\underline{p}(z')$ and $\overline{p}(z')$ for $A=6$, $M=4$, the truncated Gaussian $q(.)$ in \cite[Eq.~(12)]{csener2024-dpc-capacity} with $\sigma_x=1.8$, and $P_X=P_Z$. 

\begin{figure}[!t]
\centering
\scalebox{\figscale}{\input{pz}}
\caption{Curves of \eqref{eq:pz-bounds} for $A=6$, $M=4$, and truncated Gaussian shaping.}
\label{fig:pz}
\end{figure}

More generally, for $K \ge 1$ we obtain
\begin{align}
    \underline{p}(z') \le p(z') \le \overline{p}(z')
    \label{eq:pz-bounds}
\end{align}
for symmetric unimodal $\underline{p}(z')$ and $\overline{p}(z')$ that both converge to $p(z')$ for large $M$. Applying \eqref{eq:sum-bound2} in Lemma~\ref{lemma:symmetric-unimodal-sum} to~\eqref{eq:pyp}, we obtain the bounds $p_\textrm{min} \le p(y') \le p_\textrm{max}$ for all $y'$, where
\begin{align}
    p_\textrm{min} & = \frac{1}{A}\left(\frac{1}{d_\textrm{max}} - \frac{A}{M}\underline{p}(0)\right) \label{eq:p-bounds-2a} \\
    p_\textrm{max} & = \frac{1}{A}\left(\frac{1}{d_\textrm{min}} + \frac{A}{M}\overline{p}(0)\right) . \label{eq:p-bounds-2b}
\end{align}
The expressions \eqref{eq:p-bounds-2a}--\eqref{eq:p-bounds-2b} give
$p_\textrm{min},p_\textrm{max}\rightarrow 1/A$ for $M\rightarrow\infty$ and thus $p(y')$ becomes uniform as $M\rightarrow\infty$.
Finally, assuming $M$ is sufficiently large so $p_\textrm{min}>0$, we have
\begin{align}
    - \log p_\textrm{max} \le
    h(Y') \le - \log p_\textrm{min}
    \label{eq:plogp-bounds-2}
\end{align}
which proves that $h(Y')\rightarrow\log A$  for $M\rightarrow\infty$. This proves \cite[eq.~(15)]{csener2024-dpc-capacity} for $M\rightarrow\infty$. Finally, one may complete the proof by applying the same steps as in \cite[eq.~(18)]{csener2024-dpc-capacity}.

\section{Conclusions}
\label{sec:conc}
We showed that the DPC scheme in \cite{csener2021dirty,csener2024-dpc-capacity} can approach any rate tuple in the capacity region of a complex-alphabet MIMO broadcast channel with CSCG noise. Future work could compare the performance of the proposed scalar DPC scheme with competing methods, such as channel inversion, with concrete codes. 

\section*{Appendix}

\subsubsection*{Proof of Lemma~\ref{lemma:symmetric}}

If $f(.)$ and $g(.)$ are both symmetric then
\begin{align}
    f*g(-x) & = \int_\mathbb{R} f(y)\, g(-x-y)\, dy \nonumber \\
    & \overset{(a)}{=} \int_\mathbb{R} f(-y)\, g(x+y)\, dy \nonumber \\
    & \overset{(b)}{=} \int_\mathbb{R} f(\tilde y)\, g(x-\tilde y)\, d\tilde y = f*g(x)
\end{align}
where step $(a)$ follows by symmetry and step $(b)$ follows by substituting $\tilde y=-y$. 

\subsubsection*{Proof of Lemma~\ref{lemma:symmetric-unimodal-conv}}

Unimodality implies that $p(.)$ can have a Dirac-delta component $\delta(.)$ at $x=0$ only, i.e., we may write
\begin{align}
    p(.) = c_p \delta(.) + f(.)
    \label{eq:px}
\end{align}
for a constant $c_p$ satisfying $0\le c_p\le 1$, and for a non-negative, symmetric, unimodal $f(.)$ without $\delta(.)$ components. We may assume the derivative $f'(.)$ exists almost everywhere \cite[Prop.~2.1]{Purkayastha-98}. The $x$ where $f'(x)$ does not exist include ``jumps" in $f(.)$. For example, a negative ``jump" at $x\ge 0$ from $f(x)=a$ to $f(x+\epsilon)=b$, where $\epsilon$ is a vanishing positive number and $a>b\ge0$, becomes a $(b-a)\delta(.-x)$ component in $f'(.)$.

As in \eqref{eq:px}, consider $q(.) = c_q \delta(.) + g(.)$. We compute
\begin{align}    
    p*q(x) = c_p c_q \delta(x) + c_p g(x) + c_q f(x) + f*g(x)
    \label{eq:conv-fg}
\end{align} 
which is symmetric; see Lemma~\ref{lemma:symmetric}. The first three summands in~\eqref{eq:conv-fg} are unimodal, so it remains to show that $f*g(.)$ is unimodal. Taking the derivative for $x\ge 0$, we have
\begin{align}    
    (f*g)'(x) & = 
    \int_{-\infty}^{\infty} 
    f(\tilde y) \, g'(x - \tilde y)\, d\tilde y \nonumber \\
    & \overset{(a)}{=} \int_{0}^{\infty} 
    \left[ f(|x-y|) - f(x+y) \right]\, g'(y)\, dy
    \label{eq:symmetric-unimodal}
\end{align}
where step $(a)$ follows by substituting $y=x-\tilde y$ and because $f(.)$ and $g(.)$ are symmetric. But the term in square brackets is non-negative because $|x-y|\le x+y$ and $f(.)$ is symmetric unimodal. We also have $g'(y)\le 0$ for $y\ge 0$ because $g(.)$ is symmetric unimodal.

\subsubsection*{Proof of Lemma~\ref{lemma:symmetric-unimodal-sum}}

\begin{figure}[!t]
\centering
\scalebox{\figscale}{\input{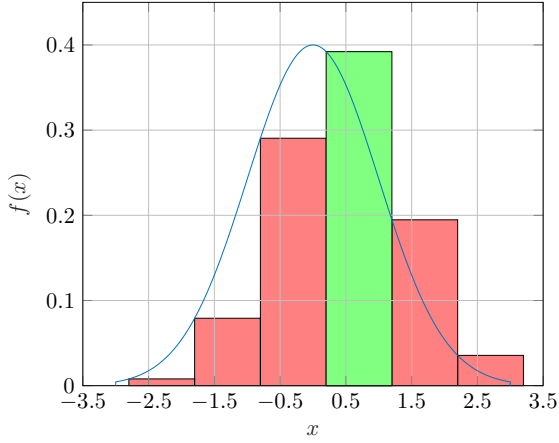}}
\caption{$f(x)$ for $A=M=6$. The sampling points are shifted by $x=0.2$ and are located at the top left corner of each of the six bars.}
\label{fig:prob}
\end{figure}

\begin{figure}[!t]
\centering
\scalebox{\figscale}{\input{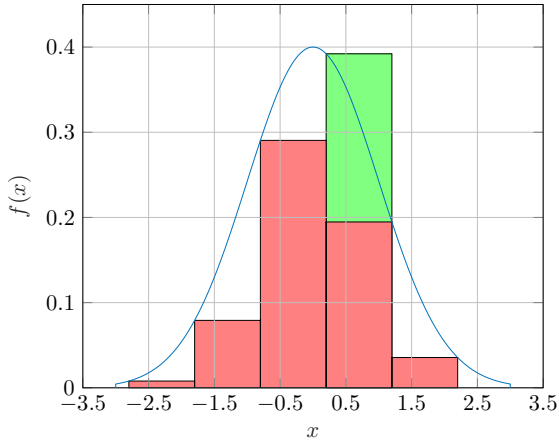}}
\caption{$f(x)$ for $A=M=6$. The area of the red bars is less than the area of $f(x)$, which is here 1.}
\label{fig:prob2}
\end{figure}

\begin{figure}[!t]
\centering
\scalebox{\figscale}{\input{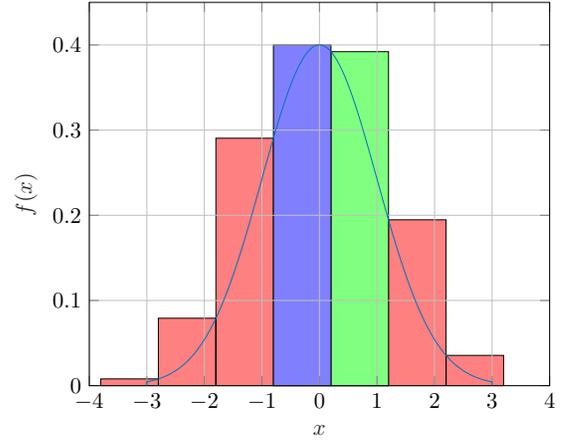}}
\caption{$f(x)$ for $A=M=6$. The area of the red, green, and blue bars is greater than the area of $f(x)$, which is here 1.}
\label{fig:prob3}
\end{figure}

The sum in \eqref{eq:sum-bound} is a Riemann sum, so we use the left and right rules of Riemann summation. Fig.~\ref{fig:prob} shows a symmetric unimodal $f(.)$ for $A=M=6$ (here $f(.)$ is a truncated Gaussian density). 
The sample points for $x=0.2$ are located on the top left corner of each of the six bars. The area of each bar is one of the summands in \eqref{eq:sum-bound}; the area of the green bar is at most $(A/M)f(0)$. Fig.~\ref{fig:prob2} shifts the red bars of positive sampling points to the left by $A/M$, and we see that the area of the five red bars is less than the integral in~\eqref{eq:sum-bound}.
This can be done for any $x$ with $0\le x<A/(2M)$, so the sum in \eqref{eq:sum-bound} is at most the integral plus $(A/M)f(0)$. Similarly, Fig.~\ref{fig:prob3} shifts the red bars of negative sampling points to the left by $A/M$ so they lie above $f(x)$.
If we add one (blue) bar of area $(A/M)f(0)$, then the sum of the areas of the seven bars is greater than the integral in \eqref{eq:sum-bound}. This proves \eqref{eq:sum-bound}.
    
To prove \eqref{eq:sum-bound2}, we perform similar steps with a countable number of bars and appropriate left and right shifts. Note that we may restrict attention $0\le x< A/(2M)$ by symmetry.

\subsubsection*{Independence of the Channel Inputs $\Xt_{k,i}$}

Let $\Xtv_{k,i}^c$ be the vector of all $\Xt_{l,m}$ except $\Xt_{k,i}$. Note from~\eqref{eq:Yk} that $\St_{k,i}$ is a function of $\Xtv_{k,i}^c$. Moreover, the chain $\Xtv_{k,i}^c- \St'_{k,i} - U_{k,i}$ is Markov because $U_{k,i}$ is chosen using $\St'_{k,i}$ as specified in \cite[Eq.~(7)]{csener2024-dpc-capacity} for $U$ given $S'$. Suppose the dithers $D_{k,i}$ are mutually independent, and consider the identities
\begin{align}
    & p(\xt_{k,i} | \xtv_{k,i}^c )
    = \int_{-A_{k,i}/2}^{A_{k,i}/2} p(d_{k,i})\, p(\xt_{k,i} | \st_{k,i},d_{k,i})\, d d_{k,i} \nonumber \\
    &\overset{(a)}{=} \sum_{u\in\mathcal{U}_{k,i}} \frac{1}{A_{k,i}}\, \underbrace{P_{U_{k,i}|\St_{k,i}'}\left(u\left|(u-\xt_{k,i})\ \textrm{mod}\ A \right.\right)}_{\textstyle \overset{(b)}{=} 2\kappa_{k,i} q(\xt_{k,i})/d(\xt_{k,i})} \nonumber \\
    &\overset{(c)}{=} p(\xt_{k,i}) 
\end{align}
where step $(a)$ follows because $\Xt_{k,i}$ is a discrete RV given $\St_{k,i}'=\st_{k,i}'$ (see~\eqref{eq:xtki}), step $(b)$ follows by the probabilistic shaping rule (see \cite[eq.~(7)]{csener2024-dpc-capacity}), and step $(c)$ follows because $2\kappa_{k,i}=A_{k,i}/M_{k,i}$ and $p(\xt_{k,i})=q(\xt_{k,i})/d(\xt_{k,i})$ (see \cite[eq.~(6) and eq.~(10)]{csener2024-dpc-capacity}). Thus, the channel inputs $\Xt_{k,i}$ in~\eqref{eq:Ytki} are mutually independent.

\subsubsection*{Entropy Bounds}
To prove \eqref{eq:entropy-bound-1}, let $\mathcal I_1=\{x: q(x)<1\}$ and $\mathcal I_2=\{x: q(x)\ge 1\}$. Observe that the length of $\mathcal I_2$ is at most one since $q(.)$ is a density. We bound
\begin{align}
    & h(X) \ge \int_{\mathbb R} - p(x) \log\frac{q(x)}{d_\textrm{min}} \,dx \nonumber \\
    & \ge \log d_\textrm{min} - \int_{\mathcal I_1} \frac{q(x)}{d_\textrm{max}} \log q(x) \,dx - \int_{\mathcal I_2} \frac{q(x)}{d_\textrm{min}} \log q(x) \,dx  \nonumber \\
    & \ge \log d_\textrm{min} + \frac{h_q(X)}{d_\textrm{max}} - \left(\frac{1}{d_\textrm{min}}-\frac{1}{d_\textrm{max}} \right) \left[q(0) \log q(0) \right]^+
\end{align}
where the final step follows for symmetric unimodal $q(.)$.
Similar steps prove \eqref{eq:entropy-bound-2} and \eqref{eq:plogp-bounds-2}.

\section{Acknowledgements}

This work was supported by the German Research Foundation (DFG) via the German-Israeli Project Cooperation (DIP) under projects KR 3517/13-1 and SH 1937/1-1.

\clearpage
\bibliographystyle{IEEEtran}
\bibliography{references}

\end{document}